\documentclass[twocolumn,showpacs,preprintnumbers,prd,superscriptaddress,nofootinbib]{revtex4}
\usepackage{epsfig}
\usepackage{graphicx}
\usepackage{amsmath,amssymb,amsfonts}
\usepackage{array}
\usepackage{url}
\usepackage{hyperref}
\usepackage{lineno}
\usepackage{xspace}

\usepackage{graphics}
\usepackage{graphicx}
\usepackage{xcolor}
\begin{document}
\def\jour#1#2#3#4{{#1} {\bf#2}, #4 (19#3)} 
\def\jou2#1#2#3#4{{#1} {\bf#2}, #4 (20#3)} 
\def\PRL{{Phys. Rev. Lett. }}
\def\EPJA{{Eur. Phys. J.  A }}
\def\EPL{{Europhys. Lett.}}
\def\PRv{{Phys. Rev. }}
\def\PRC{{Phys. Rev.  C }}
\def\PRD{{Phys. Rev.  D }}
\def\JAP{{J. Appl. Phys. }}
\def\AJP{{Am. J. Phys. }}
\def\NIMA{{Nucl. Instr. and Meth. Phys. A }}
\def\NPA{{Nucl. Phys. A }}
\def\NPB{{Nucl. Phys.  B }}
\def\NPBP{{Nucl. Phys.  B (Proc. Suppl.) }}
\def\NJP{{New J.  Phys. }}
\def\EPJC{{Eur. Phys. J.  C }}
\def\PLB{{Phys. Lett. B }}
\def\PHY{{Physics }}
\def\MPLA{{Mod. Phys. Lett. A }}
\def\PRp{{Phys. Rep. }}
\def\ZPC{{Z. Phys. C }}
\def\ZPA{{Z. Phys. A }}
\def\PPNP{{Prog. Part. Nucl. Phys. }}
\def\JPG{{J. Phys. G }}
\def\CPC{{Comput. Phys. Commun. }}
\def\APP{{Acta Physica Pol. B }}
\def\AIP{{AIP Conf. Proc. }}
\def\JHEP{{J. High Energy Phys. }}
\def\PSC{{Prog. Sci. Culture }}
\def\NCim{{Nuovo Cim. }}
\def\SNC{{Suppl. Nuovo Cim. }}
\def\SJNP{{Sov. J. Nucl. Phys. }}
\def\SPJ{{Sov. Phys. JETP }}
\def\JLet{{JETP Lett.}}
\def\PTP{{Prog. Theor. Phys. }}
\def\PTPS{{Prog. Theor. Phys. Suppl. }}
\def\IANSF{{Izv. Akad. Nauk SSSR: Ser. Fiz. }}
\def\JPCS{{J. Phys. Conf. Ser. }}
\def\AHEP{{Adv. High Energy Phys. }}
\def\IJMPE{{Int.  J. Mod. Phys. E }}
\def\ARNPS{{Ann. Rev. Nucl. Part. Sci. }}

\def\ct{\cite}
\def\sNN{\sqrt{s_{N\!N}}}
\def\sNNq{s_{N\!N}}
\def\sppq{s_{pp}}
\def\spp{\sqrt{s_{pp}}}
\def\eNN{\varepsilon_{N\!N}}
\def\pbp{{\bar p}p}

\def\col{Collab.}
\def\bi{\bibitem}
\def\ea{{\sl et al.}}
\def\eg{{\sl e.g.}}
\def\vrs{{\sl vs.}}
\def\ie{{\sl i.e.}}
\def\va{{\sl via}}
\def\nopar{\noindent}
\def\bi{\bibitem} 
\def\lssim{\stackrel{<}{_\sim}}
\def\gtsim{\stackrel{>}{_\sim}}
%


\title{
  Effective-energy universality 
  approach 
 describing
 total multiplicity 
 centrality  dependence 
 in 
 heavy-ion collisions  
}
\author{Edward K. Sarkisyan-Grinbaum}
\email {Edward.Sarkisyan-Grinbaum@cern.ch}
\affiliation{Experimental Physics Department, CERN, 1211 Geneva 23, 
Switzerland}
\affiliation{Department of Physics, The University of Texas at
Arlington, Arlington, TX 76019, USA}
\author{Aditya Nath Mishra}
\email {Aditya.Nath.Mishra@cern.ch}
\affiliation{Instituto de Ciencias Nucleares, Universidad Nacional 
Autonoma de Mexico, Mexico City, 04510, Mexico}
\author{Raghunath Sahoo}
\email{Raghunath.Sahoo@cern.ch}
\affiliation{Discipline of Physics, School of Basic Science, Indian
Institute of Technology, Indore 452020, India}
\author{Alexander S. Sakharov}
\email{Alexandre.Sakharov@cern.ch}
\affiliation{Experimental Physics Department, CERN, 1211 Geneva 23, 
Switzerland}
\affiliation{Department of Physics, New York University, New York, NY
10003, USA}
\affiliation{Physics Department, Manhattan College, Riverdale, NY 10471,
USA}

 \begin{abstract} 
 The recently proposed participant dissipating effective-energy approach  
is applied to describe
  the dependence on centrality 
 of the  multiplicity 
  of charged particles measured in heavy-ion collisions at the collision 
 energies
  up to the 
 LHC energy of 5 TeV.
 The effective-energy approach relates multihadron production in different 
types of collisions, by combining, under the proper collision energy 
scaling,
  the constituent quark picture with Landau relativistic hydrodynamics.
  The measurements are shown to be well described in terms of the 
centrality-dependent effective energy of participants 
   and an explanation of the differences in the measurements at RHIC and 
LHC are given by means of the recently introduced hypothesis of the 
energy-balanced limiting fragmentation scaling.
 A similarity between the centrality data and the data from most central 
collisions is 
 proposed 
 pointing to the central character of participant 
interactions 
 independent of 
     centrality.
  The findings complement 
  our 
 earlier 
 studies
 of the similar midrapidity pseudorapidity 
density measurements 
       extending the 
 description to the full pseudorapidity range 
 in view of 
   %
  the 
 similarity 
  of multihadron production
 in 
 nucleon interactions 
 and heavy-ion collisions. 

 \end{abstract}

\maketitle


Recently, the measurements 
of the
centrality dependence of the (mean) total multiplicity of charged 
 particles in PbPb collisions 
 at 
the center-of-mass (c.m.) energy
 $\sNN=5.02$~TeV
 have been reported 
 by
 ALICE 
\ct{alice-mult5}.
 Here, we describe these measurements in the framework of the dissipating 
energy of constituent quark participants, or, for brevity, the 
effective-energy approach, proposed by two of us in \ct{edwarda,edward}. 
Exploiting a 
concept of centrality-dependent effective energy on nucleon participants 
\ct{edward2},
 one demonstrates that the 5.02 TeV ALICE measurements can be well 
accommodated by the above approach and thus complements the results of our 
findings \ct{edward502} 
 of 
 a very good
description of the data on the measurements of the 
midrapidity pseudorapidity densities in nucleus-nucleus collisions within 
the energy range up to 5.02 TeV, provided that the the total multiplicity 
is the critical variable for obtaining the information on the 
multihadron production 
dynamics \ct{multrev,book,pprev}.  
 

 Let us give a brief description of  the effective-energy approach.
 Within this approach, 
 which    
 interrelates different types of collisions \ct{edwarda},
 the particle production process 
is quantified
 in 
 terms of the 
amount of {\it effective} energy
deposited 
into the small
 Lorentz-contracted volume 
 which is 
formed at the early stage of a 
collision.
  Then, the
 whole process of  the particle production
  is 
 considered
  as the expansion of an initial state 
  and the subsequent break-up into particles.  
This 
 picture
  resembles 
the Landau
 relativistic
 hydrodynamic model
 of 
 multiparticle production 
 \ct{Landau}.
 In the meantime, the effective-energy approach 
 considers 
the Landau 
hydrodynamics 
 being
 treated 
 in the framework of constituent (or dressed) quarks, in 
 accordance with the additive quark model \ct{additq,constitq}.
 This makes the secondary particle  production 
to be  basically driven by  the amount of the 
 initial 
 energy 
 of 
 constituent quarks 
 pumped into the Lorentz-contracted 
 overlap 
 region
 of colliding objects.
 Then, in $pp/\pbp$ collisions, a single constituent quark from  
 each nucleon is 
 assumed
  to   contribute
 in a collision.
 The remaining quarks
 are treated as spectators. 
  The spectator quarks 
 do not participate in the 
 particle 
 production, while 
 result
   into 
 formation of leading particles 
  and 
 carrying away a  significant part of the collision energy.
  Thus, 
  the effective energy for
  multiparticle production in  $pp/\pbp$ collisions is the energy of 
  a single quark pair   interaction,
  i.e. represents 1/3 of the entire nucleon energy.
 In collisions of nuclei, however, due to the large size of the 
nucleus and the long travel path inside the nucleus, more than one quark 
per nucleon can interact. 
 In 
 the most central (head-on) heavy-ion collisions, 
 where the colliding nuclei are almost fully overlapped, 
 all three constituent quarks from each of the participating nucleons 
 may interact 
  and deposit their energy 
 into the collision zone.
  Then
the whole energy of the 
 nucleons 
 becomes available for the 
 particle production.  
  Within this picture,
 the 
 bulk 
 measurements
 in 
 head-on heavy-ion collisions 
 at 
 $\sqrt{s_{NN}}$
 are 
 expected 
 to be similar to 
  those from
 $pp/\pbp$ collisions 
but at 
  a
 three times larger 
 c.m. energy $\sqrt{s_{pp}}$, 
  i.e. 
 at 
 $\sqrt{s_{pp}} \simeq 
 3\sqrt{s_{NN}}$.
 Let us stress here that the effective-energy approach is considered 
being applied to the bulk variables, while it is well understood that 
collective effects such as elliptic flow, correlations would provide 
furter important details but to be addressed in separate studies.

 Combining 
 within the above consideration
 the 
 two contributing 
 ingredients, namely the 
 constituent quark picture and  the pseudorapidity density 
 from the Landau hydrodynamics, 
  one obtains 
 the relationship between 
 the 
charged particle rapidity density
per participant pair, 
 $\rho(\eta)=(2/N_{\rm{part}})dN_{\rm{ch}}/d\eta$, 
in heavy-ion 
 collisions  and in  $pp/\pbp$  
 interactions \ct{edwardPRD}:

 \begin{eqnarray}
 &\!\!\!\!\! &    
 \nonumber
\frac{\rho(\eta)}{\rho_{pp}(\eta)}\!=\!
\frac{2N_{\rm{ch}}}{N_{\rm{part}}\, N_{\rm{ch}}^{pp}}
\, \sqrt{\frac{L_{pp}}{L_{N\!N}}} \,
 \exp\! \left[ \frac{\eta^2}{2} \left(
 \frac{1}{L_{pp}}
 - 
 \frac{1}{L_{N\!N}} 
 \right) \right] \, , \\
 &\!\!\!\!\! &    
 \spp=3\sNN\, .
 \label{rapdistf}
 \end{eqnarray}
 %
 %
 Here, 
 $N_{\rm{ch}}$ and 
 $N_{\rm{ch}}^{pp}$ are the (total) mean multiplicities in nucleus-nucleus 
and nucleon-nucleon collisions, respectively, and $N_{\rm {part}}$ is the 
number of nucleon 
participants. The relation of the pseudorapidity density 
and the 
mean multiplicity is applied in its Gaussian form as obtained in Landau 
hydrodynamics. The factor $L$, which is related to the Lorentz contraction of the system,
 is defined as $L = {\ln}({\sqrt{s}}/{2m})$. 
According to the approach considered, $m$ is the proton mass, $m_{p}$, in 
nucleus-nucleus collisions and
 represents the constituent quark mass ($m_{q}$) in $pp/\pbp$ collisions 
 set
 to  $\frac{1}{3}m_{{p}}$.


 At the midrapidity, $\eta\! \approx\! 0$,  Eq.~(\ref{rapdistf}) 
simplifies  to: 

\begin{eqnarray}
\nonumber
\frac{\rho(0)}{\rho_{pp}(0)}&=&
\frac{2N_{\rm{ch}}}{N_{\rm{part}}\, N_{\rm{ch}}^{pp}} 
\sqrt{\frac{L_{pp}}{L_{N\!N}}} \ , \\
\sNN&=&\spp/3 \, .
 \label{eqn2}
\end{eqnarray}
Taking into account that $L_{pp} = \ln \left( \spp/2m_q \right)$ 
 and  $L_{N\!N} = \ln \left( \sNN/2m_p \right)$ and setting
  $m_q=m_p/3$ 
 and  $\spp=3\sNN$, one gets for Eq.~(\ref{eqn2}):

\begin{eqnarray}
\nonumber
\frac{\rho(0)}{\rho_{pp}(0)}&=&
\frac{2N_{\rm ch}}{N_{\rm part}\, N_{\rm ch}^{pp}} 
\sqrt{1 -
\frac{4\, \ln 3}{\ln \left( 4m_p^2/s_{NN}\right) }} \ ,  \\ 
\sNN&=&\spp/3 \, .
\label{eqn3}
\end{eqnarray}
\nopar

Solving Eq.~(\ref{eqn3}), for the multiplicity $N_{\rm{ch}}$ at a given midrapidity density  pseudorapidity $\rho(\eta\! \approx\! 0)$ at $\sqrt{s_{NN}}$, and for the rapidity density  $\rho_{pp}(0)$ and the multiplicity $N_{\rm{ch}}^{pp}$ at $3 \sqrt{s_{NN}}$, one finds:

\begin{eqnarray}
\nonumber
\frac{2N_{\rm{ch}}}{N_{\rm{part}}}  &=&
N^{pp}_{\rm ch} \,
\frac{\rho(0)}{\rho_{pp}(0)} \,
\sqrt{1-\frac{2 \ln 3}{\ln\, (4.5 \sNN/m_ p)}}\ ,  \\
	    \sNN&=&\spp/3 \, .
\label{eqn4}
\end{eqnarray}
\nopar

 In the 
 further development  \ct{edward2}, 
 one 
 considers 
 this dependence 
 in
terms of centrality.
 The centrality 
 is closely 
related to the number of nucleon participants
 determined using Monte Carlo Glauber calculations, 
  so that 
  the largest number of participants 
  contribute to the most central heavy-ion collisions. 
 In the meantime, the centrality is 
 regarded as the degree  of the overlap of the  volumes of the 
 two colliding  nuclei, 
 characterized  by the impact parameter.
 The smaller the impact parameter (more central collisions), the
larger is  the overlap zone.
 Considering the 
 volumes of the colliding nuclei projected onto the
overlapped area 
 as 
being populated by the number of 
participants depositing their energy into the 
Lorentz-contracted volume of the early stage collision zone, the 
centrality, in the participant effective-energy approach, can 
  be  
 related to the amount of the energy 
 released
 for 
 particle production, 
 i.e.   to the effective energy, $\eNN$, of  
the participants.
 Then
 this effective energy 
 can be 
defined as 
a fraction of the c.m. energy available in a collision according to the 
centrality, $\alpha$:
 %
 \begin{equation}
\eNN = \sqrt{s_{NN}}(1 - \alpha).
\label{Eeff}
\end{equation}
  One has to note that this relation is satisfied for all but most 
peripheral collisions, where no overlap zone can be actually defined 
and the collisions rather resemble diffractive interactions of 
nucleons, 
so no scale factor 1/3 to be applied anymore. 

  Conventionally, the data are divided into 
 classes of centrality, or 
  centrality intervals, so that $\alpha$ is the average 
 centrality per 
 centrality interval, \eg\ $\alpha = 0.25$ for the centrality 
  interval of 
 20--30\% centrality.

Then, for non-central collisions, 
 Eq.~(\ref{eqn4})  
 reads
 
\begin{eqnarray}
& &
\nonumber
\frac{2N_{\rm{ch}}}{N_{\rm{part}}}  =
  N^{pp}_{\rm ch} \,
   \frac{\rho(0)}{\rho_{pp}(0)} \,
  \sqrt{1-\frac{2 \ln 3}{\ln\, (4.5\ \eNN /m_ p)}}\,,  \\
& &    \eNN=\spp/3 \, .
  \label{pmultc}
\end{eqnarray}
  \nopar
 considering
 central 
collisions of nuclei
  at 
 the effective c.m.  energy 
 $\eNN$. Here
 $\rho_{pp}(0)$ 
 and 
 $N_{\rm{ch}}^{pp}$  
 are
 taken 
 from  $pp/{\bar p}p$ 
 data
 at  $\spp=3\, \eNN$.

Let us note that each of the scalings introduced by  Eqs.~(\ref{eqn4}) and 
~(\ref{Eeff}) 
regulates a particular physics ingredient 
 of the
 modelling within the 
participant dissipative effective-energy  
approach. The scaling embedded in Eq.~(\ref{eqn4}) reflects the 
constituent 
quark picture and then reveals a similarity of multihadron 
production in hadronic and 
nuclear collisions. The scaling driven by Eq.~(\ref{Eeff}) addresses the 
energy budget effectively retained in the most central collisions while 
defining the energy availability for the global variables in non-central 
collisions.


\begin{figure*}
 \begin{center}
\resizebox{0.65\textwidth}{!}{%
  \includegraphics{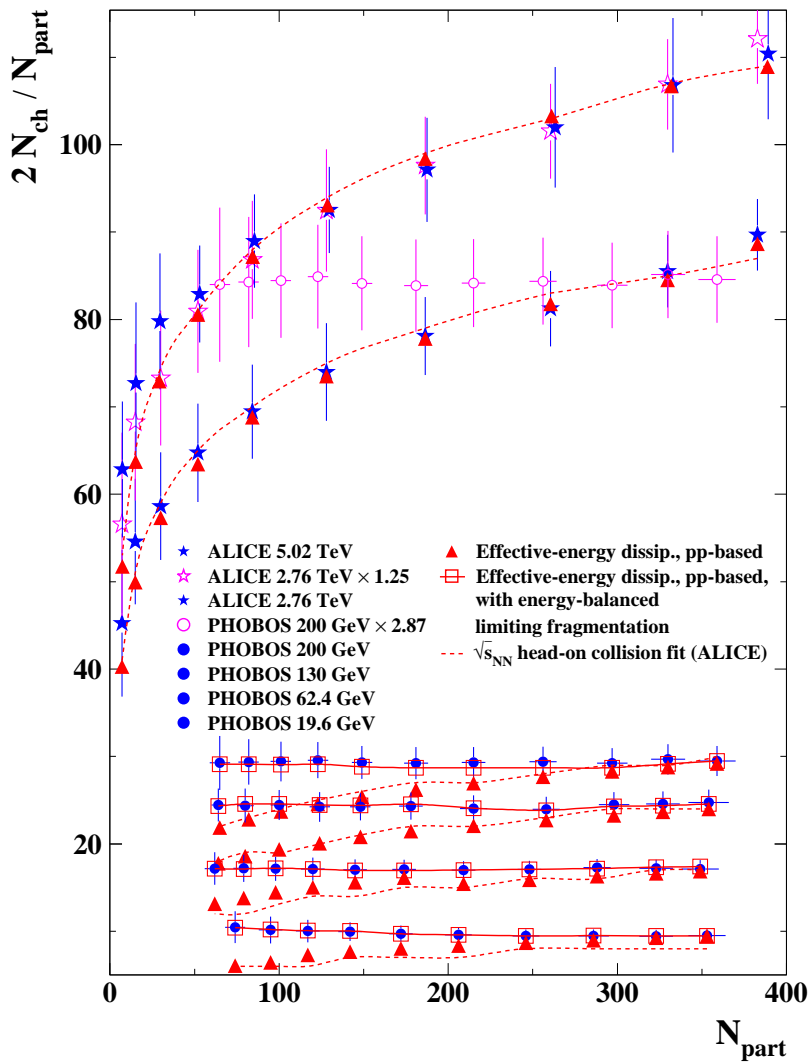}}
 \end{center}
\caption{
The
 charged particle
 mean multiplicity
per participant
pair as a function of the number of participants, $N_{\rm part}$.
 The
solid
 stars show the dependence measured in 
PbPb collisions at the LHC by the ALICE experiment at $\sNN=2.76$~TeV
\ct{alice-mult2} and 5.02 TeV \ct{alice-mult5}, 
and the
 solid
 circles show the 
 measurements from 
 AuAu collisions at RHIC  by
the PHOBOS experiment at $\sNN=19.6, 62.4, 130$ and 200~GeV
\ct{phobos-all} (the symbols indicate $\sNN$ 
increasing
bottom to top). 
 The
 triangles show the calculations
by Eq.~(\ref{pmultc}) using $pp/\pbp$ data within the participant 
dissipating effective-energy approach.
 The
 open
 squares show
the effective-energy
  calculations
 which include
 the energy-balanced
limiting fragmentation scaling
 (see
text);
the solid lines connect the calculations
 to guide the eye.
 The  dashed  lines represent
the
 calculations
 using the ALICE fit \ct{alice-mult5}
 to the c.m. energy
dependence of the
 mean multiplicity in the most central heavy-ion collisions.
 The
 open
 stars show the ALICE measurements at $\sNN=2.76$~TeV
multiplied by
1.25, and 
 the
 open
 circles show the PHOBOS measurements at $\sNN=200$~GeV
multiplied by
2.87.
 }
\label{Fig1}       
\end{figure*}
%

 Figure~\ref{Fig1} shows 
 the effective-energy calculations by Eq.~(\ref{pmultc}) 
 compared to 
 the charged particle multiplicity, $N_{\rm 
ch}/(N_{\rm part}/2)$,
  as a function of the number of participants $N_{\rm part}$
   as  measured in heavy-ion collisions at $\sNN$ of TeV energies 
 by the ALICE  experiment at the LHC \ct{alice-mult2,alice-mult5}
 and, at GeV energies, by the PHOBOS experiment at RHIC \ct{phobos-all}. 
 In the calculations, the midrapidity density $\rho_{pp}(0)$ and the 
multiplicity 
$N_{\rm{ch}}^{pp}$ are taken from the existing $pp/\pbp$ data 
\ct{pprev,pdg18}, and the $\rho(0)$ values are taken from the heavy-ion 
collision data \ct{phobos-all,alice276c,alice502c} 
 where 
 available. 
   Where no 
 data 
  exist, 
 the  corresponding experimental c.m. energy fits
 are  used.
 The 
 linear-log \ct{pprev} 
 and 
 power-law \ct{alicepp8} 
 $s_{pp}$-fits 
 for  
 $\rho_{pp}(0)$ at 
$\sqrt{s_{pp}}\leq$~53~GeV
 and 
at  
$\sqrt{s_{pp}}>$~53~GeV, 
respectively, along with 
 the 
 power law c.m. energy fits 
 for $N_{\rm{ch}}^{pp}$ \ct{edwardPRD}
 and 
  $\rho(0)$ \ct{alice502c}
  are used.
 
 One can see from Fig.~\ref{Fig1} that within the dissipating participants 
effective-energy 
 approach,
 where the collisions are drived by the centrality-defined effective c.m. 
 energy $\eNN$, the 
 calculations well reproduce   
  the 
 centrality dependence obtained in the TeV-energy region from LHC,
 slightly underestimating a couple of the most peripheral 
 measurements.  
 However,
 for the RHIC data,
  the deviation between the   
  measurements and the calculations is seen already 
 for middle $N_{\rm part}$ values.
 The difference in the behaviour of the data obtained at RHIC and at LHC 
becomes more clearer as soon as one multiplies the RHIC 200 GeV data by a 
factor 2.87 in order to match the ALICE 2.76~TeV  data from highly central 
collisions. 
 In the meantime, one can observe 
 that there is almost no difference between the 2.76 TeV and 
 5.02 TeV
 LHC data,
 where the lower-energy 
 measurements are 
multiplied by 1.25 to match the higher-energy ones.
 
 The differences 
 observed have been discussed in 
\ct{edwardPRD} and an explanation has been given by introducing the 
energy-balanced 
limiting 
fragmentation scaling for the pseudorapidity spectra in 
non-central collisions. By means of this scaling, the pseudorapidity 
distributions 
in 
heavy-ion collisions
at RHIC energies
 are shown to be reproduced 
 resulting into the 
 centrality independence of the 
 multiplicity,
 see Fig.~\ref{Fig1}.

 In what follows,
 the energy-balanced limiting fragmentation hypothesis is applied as it is 
 elaborated 
in  \ct{edwardPRD}. 

 First, let us notice that,
 as it is outlined above, in the picture of the 
effective energy 
 approach, the global observables are defined by the energy of the 
 participating constituent quarks pumped into the overlapped zone of the 
 colliding nuclei. Hence, the bulk production is driven by the initial 
 energy deposited at zero time at rapidity $\eta=0$, similar to the
 Landau hydrodynamics. Then, as
  is expected and indeed found in \ct{edward502,edward,edward2,edwarda}, 
  pseudorapidity density (and pseudorapidity transverse energy density) 
  {\it 
  at midrapidity} is well reproduced for {\it all} centralities and all 
available energies.

Meantime, the data shown in Fig.~\ref{Fig1}, represent the 
{\it total} multiplicities, i.e. addresses  Eq.~(\ref{rapdistf}) 
after its integration over the {\it full}-$\eta$ spectrum. 
 Then, the fragmentation regions contribute, in addition to the 
midrapidity region, and this contribution has to be taken into account.
 This point has been addressed in  \ct{edwardPRD}, where the rapidity 
spectra were studied. 

 As it was found in \ct{edwardPRD}, the calculations well reproduce the 
data on the full pseudorapidity spectra at all c.m. energies as soon as 
the effective-energy approach is applied to the most central collisions. 
As a consequence, the total multiplicity values for central collisions are 
well reproduced, as one can indeed see in Fig.~\ref{Fig1}, the points at 
the large $N_{\rm part}$. This is understood as soon as in the 
calculations, the energy is considered to be deposited into the overlapped 
zone and then, $\sNN$ is one driving the particle-production process in 
the most 
central collisions. Consequently, in Eq.~(\ref{rapdistf}) the values of 
the contributed variables are taken from the most central collisions. 

For non-central collisions, as discussed above, 
not the full c.m. energy  is considered to contribute to the 
particle production but   
the effective energy $\eNN$ to be assumed instead. 
 As soon as this has been applied to the calculations, the calculated 
pseudorapidity spectra were found \ct{edwardPRD} to be narrower than the 
measured ones for the pre-LHC data, so that the fragmentation region were 
not well reproduced. Consequently, 
 as shown in Fig.~\ref{Fig1},
the non-central values, measured at RHIC, 
are seen to be higher and to follow surprisingly  constant values in 
disagreement with the lower, monotonically increasing   
calculations and, interestingly, with the LHC data, to which the 
calculations are found well agree.
 The  deviation between the calculations and the pre-LHC 
measurements is not surprising and can be explained due to a smaller value 
of $\eNN$, used in the calculations, compared to the value of the actual 
collision energy $\sNN$, as well as due to the fact that the calculations,
 similar to the Landau approach, are undertaken in the assumption of the
head-on character of nuclei collisions, which is clearly not the case for 
non-central collisions.
 Therefore, the pseudorapidity distributions have to be corrected in 
order to balance the energy used in the calculations vs. that in the 
measurements, to account for the fragmentation region.
 These, as discussed below and in \ct{edwardPRD}, seem also to explain the 
difference between the LHC and pre-LHC total multiplicity measurements.

   To address the point of 
 the fragmentation region, let us first recall the 
hypothesis 
of the limiting fragmentation scaling 
 \ct{limfrag}, which 
 states
   that at high enough energies, the (pseudo)rapidity density spectra for 
 given interacting particle types become 
   independent of the c.m. energy in the fragmentation region when shifted 
by 
the beam rapidity $y_{\rm beam}
    = \ln(\sNN/m_p)$: $\eta\to \eta -y_{\rm beam}$. 

   As soon as, in the effective-energy approach, 
 the particle 
production is considered to be 
 drived by the energy of the participants involved in the overlapped zone,  
 one would 
 naturally 
 expect that the  
behaviour similar to that of the limiting fragmentation 
 to hold for the calculated pseudorapidity distribution but when the 
latter  is shifted 
by the rapidity defined by the effective energy $\eNN$, namely
 $y_{\rm eff}= \ln(\eNN/m_p)$. 
 Indeed,  in \ct{edwardPRD}, 
it was 
 found that 
  for a non-central collision, the effective-energy calculation of the  
pseudorapidity distribution 
 matches immediately the measured spectra as soon as the former  is 
shifted by $y_{\rm eff}$ while the latter  by 
    $y_{\rm beam}$. 
 This effect  is observed 
 to hold independently of centrality, as 
 soon as 
the  
corresponding $y_{\rm eff}$ shift 
 was applied.  

 As so, this 
 unambiguously points out that, in order  to describe the unshifted 
$\eta$-distribution, 
  one 
    needs to take into account the difference between the c.m. 
energy in the measurements and the effective energy of the participants 
in the calculations, 
 i.e.
 in the other words  to {\it balance} the energy. 
 To this end, 
the calculated distribution, 
Eq.~(\ref{rapdistf}), is shifted by 
the 
energy difference, $y_{\rm eff}-y_{\rm beam}=\ln (\eNN/\sNN$), in the 
fragmentation 
region, or, due to Eq.~(\ref{Eeff}), by $\ln (1-\alpha)$. 
  This immediately put in agreement    
the calculated distribution and the  
measurements for non-central collisions with no deficit in the 
fragmentation region  \ct{edwardPRD}.
 Due to this, the  hypothesis was named 
 the 
  energy-balanced limiting fragmentation scaling. 
 Using this scaling, 
 the  calculated pseudorapidity density spectrum 
  gets 
  corrected outside the central-$\eta$ region for all centralities.
   It is clear that in head-on a collisions, $\alpha$ tends 
   to zero
    what
    makes the shift negligible. 

 To add is that where no  pseudorapidity density distributions are 
available in $pp/\pbp$ data
 at
$\spp= 3\, \eNN$, and therefore no integration is possible using 
Eq.~(\ref{rapdistf}), the energy-balanced limiting fragmentation scaling 
is
 applied to reproduce the calculated $\rho(\eta)$.
 The measured
distribution from  a
non-central
heavy-ion collision
is shifted by $(y_{\rm beam}-y_{\rm eff})$,
 i.e. 
 $\eta\to
 \eta+\ln(1-\alpha)$.
 Then
 $N_{\rm ch}$ is calculated by adding
 the difference between the integral from
the obtained shifted
 distribution and the measured multiplicity to the midrapidity calculation 
of
  Eq.~(\ref{pmultc}).
 Note that for central (midrapidity) region, there is no need to apply 
the energy-balanced limiting fragmentation, as soon as particle production 
in this region is well described by the effective energy approach of 
centrally colliding participants, as it is discussed above.  

 Using
 this
 ansatz,
 the values of $N_{\rm ch}$, calculated for
 each
 centrality for the RHIC measurements,
 found to
reproduce well the pre-LHC data, 
 as 
    shown 
 in Fig.~\ref{Fig1}. 

 In the meantime,
 one can also 
  conclude 
from Fig.~\ref{Fig1} 
 that, in contrast to the pre-LHC energies,
 almost no energy-balanced contribution
 is needed  for the
 calculations 
to
describe the
LHC mean multiplicity measurements at TeV c.m. energies.
  Given this observation and the fact that   
 the calculations imply they are made 
 in an assumption of central nucleus-nucleus collisions at the c.m. energy 
of 
 the value of 
 $\eNN$, 
 one
 can conclude that
 in TeV-energy
  heavy-ion collisions,
    the multihadron production obeys a head-on collision regime,
 for all the
 centralities
 measured. 
  This conclusion is supported 
 by the one made in 
\cite{edwardPRD} where the 
$\eNN$ dependence of the  centrality multiplicity data are shown to 
well 
follow 
and so complementing the $\sNN$ 
dependence of the  head-on multiplicity data without taking 
into 
account the energy balance.
This feature is also clearly seen here,  
 as shown in 
 Fig.~\ref{Fig1}, 
 where 
 the 
 ALICE  heavy-ion
 {\it head-on}  $\sNN$ fit 
  \ct{alice-mult5} is 
 demonstrated
to well follow the LHC {\it centrality} data 
expressed in terms of $\eNN$.

  The above findings may be treated  pointing to  apparently different 
regime of
hadroproduction
  occurring
 in heavy-ion
collisions
 as $\sNN$
 moves from 
 GeV 
 to TeV energies.
 This conclusion finds its support 
 also 
 in 
 our results  \ct{edward502}
 of 
 describing the 
 midrapidity density  $\sNN$-dependence as soon as the LHC data are 
included.

 The energy-balanced limiting fragmentation provides also 
an explanation  
 to the unexpected difference of the midrapidity pseudorapidity density 
increase 
\vrs\ the total multiplicity independence with the increase of the 
number of 
nucleon participants as measured at RHIC (RHIC ``puzzle'' 
\cite{edwardPRD}).
At GeV energies,  the multiplicity gets an additional fraction of the 
energy as the difference between the collision c.m. energy shared by all 
nucleons and the effective energy of the participants driving the 
particle production process. No such difference appears at the LHC where 
both variables are found to increase with the number of 
participants, in agreement with the central character of all-centrality 
collisions as discussed just above.






 Let us note in conclusion that
 the picture of the effective-energy approach is shown as well
  to 
 explain \ct{edwarda,edward}
 the similarity of the measurements in other collisions, such the 
 scaling between the charged particle mean multiplicity in $e^+e^-$ and 
$pp/\pbp$ collisions \ct{lep} and the universality of both the 
multiplicity and the midrapidity density measured in the most central 
nuclear collisions and in $e^+e^-$ annihilation \ct{phobos-sim}; see 
\ct{book,pprev} for discussion.
 In the latter case, colliding leptons are considered to be structureless 
 and deposit their total energy into the Lorentz-contracted volume. 
 This 
is
 shown \ct{pprev} to be supported 
 by 
the observation 
 that
 the multiplicity and $\eta$-distributions 
 in $pp/\pbp$ interactions 
 are
 well reproduced by $e^+e^-$ data as soon as the inelasticity 
is set to $\approx$0.35, \ie\ effectively
  1/3 of the  hadronic interaction energy.
  For recent discussion on the universality of hadroproduction up to LHC 
energies, see \ct{pdg18}.

  Summarizing,
 the effective-energy dissipation approach based on the picture 
 combining 
 the constituent quark model together with Landau relativistic 
hydrodynamics in 
 sense
  of the universality of the multihadron production in 
hadronic and nuclear collisions is shown to well describe the total 
 (mean) multiplicity data 
 in
 GeV to TeV c.m. energy   
 heavy-ion collisions. 
 In particular, 
 the centrality dependence of 
 the multiplicity 
 of charged particles measured 
 up to 5 
 TeV are shown to be well reproduced.
 In addition,
 the centrality dependence of the data is shown to be 
 in good agreement with the fit to the head-on c.m. energy dependence 
 confirming the central character of the collisions independent of the 
 centrality measured. The observation of differences between the RHIC and 
 the LHC measurements suggests the change of the particle 
 production regime  as the c.m. energy of heavy-ion collisions
moves 
from GeV to TeV energy region.
 This study complements and supports the results from our 
 earlier 
 investigations \ct{edward502} of the midrapidity pseudorapidity density 
centrality 
dependence measurements in the same energy
 range while now made 
 for the full-rapidity interval.
 Foreseen measurements at LHC and future colliders at higher 
energies and 
 with 
 different  
 types of colliding objects 
 are 
   of 
 high interest 
 in  
 clarifying the features of the 
participant 
 effective-energy approach 
  and 
 in  
 view of 
 better understanding of the mechanism
 of the hadroproduction process.
 \\
 

 Raghunath Sahoo acknowledges the financial support from
Project No. SR/MF/PS-01/2014-IITI(G) of Department of Science \& 
Technology, Government of India.
  The work of Alexander Sakharov is partially supported by the US National 
Science Foundation under Grants No. PHY-1505463 and No. PHY-1402964.

\end{document}